\documentclass[superscriptaddress,twocolumn,showpacs,a4paper,
amssymb,amsmath,nobibnotes,aps,prl,showkeys,nofootinbib,notitlepage]{revtex4-1}
\usepackage{verbatim}
\usepackage[T1]{fontenc}
\usepackage[utf8]{inputenc}
\usepackage[american]{babel}
\usepackage{epsfig}
\usepackage{booktabs}
\usepackage{multirow}
\usepackage{dcolumn}
\usepackage{amsmath}
\usepackage{mathtools}
\usepackage{amsfonts}
\usepackage{amssymb}
\usepackage{ulem}
\usepackage{epstopdf}
\usepackage{bm}
\usepackage{siunitx}
\usepackage{braket}
\usepackage{enumitem}
\usepackage{soul}
\usepackage[table]{xcolor}
\usepackage{color}
\usepackage{transparent}
\usepackage{pifont}
\usepackage{makecell}
\usepackage{booktabs}
\usepackage{multirow}
\usepackage{dcolumn}
\usepackage{colortbl}
\usepackage{orcidlink}

\usepackage{hyperref}
\hypersetup{
    colorlinks=true,     linkcolor=royalblue, 
    citecolor=magenta}

\definecolor{navyblue}{rgb}{0.0, 0.0, 0.5}
\definecolor{royalblue}{rgb}{0.25, 0.41, 0.88}
\definecolor{cadmiumgreen}{rgb}{0.0, 0.42, 0.24}
\definecolor{blue-violet}{rgb}{0.54, 0.17, 0.89}
\definecolor{darkviolet}{rgb}{0.58, 0.0, 0.83}
\definecolor{orange(colorwheel)}{rgb}{1.0, 0.5, 0.0}

\newcommand{\udt}[3]{#1^{#2}_{\phantom{#2}#3}}

\begin{document}

\title{$f(T)$ gravity after DESI Baryon Acoustic Oscillation and DES Supernovae 2024 data}

\author{Celia Escamilla-Rivera\orcidlink{0000-0002-8929-250X}}
\email{celia.escamilla@nucleares.unam.mx}
\affiliation{Instituto de Ciencias Nucleares, Universidad Nacional Aut\'{o}noma de M\'{e}xico, 
Circuito Exterior C.U., A.P. 70-543, M\'exico D.F. 04510, M\'{e}xico.}

\author{Rodrigo Sandoval-Orozco}
\email{rodrigo.sandoval@correo.nucleares.unam.mx}
\affiliation{Instituto de Ciencias Nucleares, Universidad Nacional Aut\'{o}noma de M\'{e}xico, 
Circuito Exterior C.U., A.P. 70-543, M\'exico D.F. 04510, M\'{e}xico.}


\begin{abstract}
\noindent
In this work, we investigate new constraints on $f(T)$ gravity using the recent Baryon Acoustic Oscillation (BAO) data released by the Dark Energy Spectroscopic Instrument (DESI) and the Type Ia supernovae (SNIa) catalog from the full 5-years of the Dark Energy Survey Supernova Program (DES-SN5YR). The $f(T)$ cosmological models considered are characterised by power law late-time accelerated expansion. Our results show that the combination DESI BAO +$r_d$ CMB Planck suggests a Bayesian preference for late-time $f(T)$ cosmological models over $\Lambda$CDM, obtaining a value of $H_0= 68.3^{+3.0}_{-3.5}$[km/s/Mpc] in agreement with SH0ES collaboration, however, due to a bigger uncertainty. 
\end{abstract}

\maketitle


The Hubble tension has reached a statistical significance of $\sim$5$\sigma$, 
strongly proven a mismatch between the cosmic late-time expansion rate $H_0$ measured through the local distance ladder method using Type Ia supernovae measurements \cite{Riess:2021jrx,Breuval:2024lsv} $H_0=73\pm1$ [km/s/Mpc], and the inferred $H_0$ value from observations of the Cosmic Microwave Background (CMB) radiation, $H_0=67.4\pm0.5$ [km/s/Mpc] \cite{Planck:2018vyg}. While cautious studies on a possible systematic origin of this mismatch have been performed by the SH0ES collaboration \cite{Riess:2021jrx}, there is no signal that this could be the reason for this $H_0$ tension issue. This result has brought interesting opportunities to change the view on how the standard cosmological model is designed, allowing us to introduce a path beyond the $\Lambda$-Cold Dark Matter(CDM) model.

Current BAO measurements released by the \textit{Dark Energy Spectroscopic Instrument} (DESI) \cite{DESI:2024uvr,DESI:2024lzq} seem to point towards new physics in the dark energy cosmological scheme~\cite{desicollaboration2024desi}. Additionally, the \textit{Dark Energy Survey 5-year SNIa} (DES-SN5YR) release \cite{DES:2024tys}, suggest a $2\sigma$ preference for a time-varying dark energy. More discussions on this aspect have been presented in \cite{Yin:2024hba,Cortes:2024lgw}.
As part of the efforts to find well-constrained proposals with these releases and also, that can address the cosmological tensions, some studies have been developing in these short period, e.g. new constraints on axion-early dark energy model \cite{Qu:2024lpx} which are not tighten even after the inclusion of DESI BAO data, interacting dark energy models \cite{Wang:2024hks,Giare:2024smz} which shows a value of $H_0$ in good agreement with SH0ES collaboration, for quintessence scalar field model \cite{Berghaus:2024kra,Tada:2024znt} showing a preference within 2-4\% for a kinetic scalar field energy, for dark energy models inspired in thermodynamics and parametrised equation-of-state in Taylor expansions \cite{Carloni:2024zpl,wang2024dark}, the first one finding that log-corrected parameterisation could alleviate the $H_0$ tension, and finally Gaussian reconstructions on quintom modified cosmology \cite{Yang:2024kdo}.
All these studies aim to increase the value of $H_0$ inferred. On one hand, the main challenge in the early CMB measurements seems to be settled in computing the angular scale of the CMB acoustic peaks ~\cite{Planck:2018vyg}. Therefore, increasing the value of $H_0$ without modifying the acoustic scale requires a different post-recombination epoch \cite{Knox:2019rjx}. On the other hand, late-time cosmic proposals require new physics that can change cosmic distances to compensate for the higher value of $H_0$, taking into account the preservation of the CMB history. 

Within these efforts, \textit{extended theories of gravity} have been proposed as a good description of a fundamental theory of gravity that allows addressing theoretical and observational issues with viable solutions in the observed mismatch  \cite{Bahamonde:2021gfp,Aguilar:2024cga,Briffa:2023ern,Sandoval-Orozco:2023pit,Nunes:2018evm,Kumar:2022nvf,Nunes:2018xbm,dosSantos:2021owt,Hashim:2021pkq}. To formulate an extension to General Relativity, we consider a construction through the metric-affine gravity \cite{BeltranJimenez:2019esp}, where \textit{teleparallel gravity} (TG) has a curvature-free connection \cite{Bahamonde:2021gfp,Krssak:2018ywd} with a scenario that include a \textit{teleparallel equivalent of general relativity} (TEGR). This theory has described a set of field equations which are dynamically equivalent to the GR ones. Within this scheme, $f(T)$ gravity emerges as a generalisation of the TEGR Lagrangian with a function of the torsion $T$ as $f(T) = -T + \mathcal{F}(T)$. 

In this work, we show that the new DESI BAO plus $r_d$ CMB Planck data release gives a slight Bayesian preference for extended $f(T)$ cosmologies, despite the fact that the parameters are within 2$\sigma$ confidence level (C.L) from $\Lambda$CDM. In addition, it is noted that high/low-z observations could be better explained in these models in comparison to $\Lambda$CDM and with first principle reasons. In such a scheme, we also consider baseline with DES-SN5YR, which gives a lower value of $H_0$ for this kind of supernovae catalog. 


\begin{table*}[htpb!]
\begin{center}
\renewcommand{\arraystretch}{0.6}
\resizebox{\textwidth}{!}{
\begin{tabular}{l |c |c |c |c |c |c c c c c c c c c c }
\hline
\textbf{Parameter} & \textbf{DESI+BBN} & \textbf{DESI+CMB\footnote{CMB Distance priors.}} & \textbf{DESI+CC } & \textbf{\thead{DESI+CMB\\{+CC+$\text{PN}^{+}$}}} & \textbf{\thead{DESI+CC\\+PN$^{+}$}} & \textbf{\thead{DESI+CC\\+SNYR5}}  \\ 
\hline\hline 
$ H_0  $ [km/s/Mpc]& $ 68.8^{+1.3}_{-1.2} $  & $ 68.44^{+0.87}_{-0.83}  $ & $ 70.6^{+6.8}_{-6.7} $ & $68.80^{+0.81}_{-0.84} $ & $72.8^{+2.1}_{-2.0}$  & $66.3^{+6.2}_{-6.1}$ \\  
$ \Omega_{\mathrm{cdm}}  $ & $ 0.239^{+0.027}_{-0.025}  $ & $ 0.2382^{+0.0095}_{-0.0097} $ & $ 0.238^{+0.031}_{-0.030}$ & $0.2349^{+0.0085}_{-0.0089}$ & $0.254^{+0.023}_{-0.022}$ & $0.309^{+0.023}_{-0.022}$ \\ 
$ w_b  $ & $ 0.02218\pm0.00077 $ & $ 0.02259^{+0.00034}_{-0.00033} $ & $ 0.0250^{+0.011}_{-0.0099}  $ & $ 0.02271^{+0.00035}_{-0.00034} $ & $0.0282^{+0.0038}_{-0.0036}$ & $0.0199^{+0.0081}_{-0.0079}$ \\
$ \Omega_\mathrm{m} $ & $ 0.286^{+0.027}_{-0.025}  $ & $ 0.286^{+0.011}_{-0.011} $ & $ 0.288^{+0.027}_{-0.025} $ & $0.2829^{+0.0095}_{-0.0099}$  & $0.308^{+0.022}_{-0.021}$ & $0.353^{+0.020}_{-0.019} $ \\ 
$ r_\mathrm{d}  $ [Mpc] & $ 149.1^{+3.5}_{-3.4} $ & $ 149.06^{+0.73}_{-0.71}  $ & $ 146.0\pm 14.0 $ & $149.03\pm 0.73$ & $148.6^{+4.6}_{-4.7} $ & $147^{+15}_{-14} $  \\
$ M $ & $ - $ & $ - $ & $ -$ & $-19.391^{+0.027}_{-0.033} $ & $-19.265^{+0.056}_{-0.059}$ & $-19.53^{+0.19}_{-0.21} $ \\ 
\hline\hline
$\chi^2_{\text{min}}$ & $8.92$ & $12.47$ & $23.99$ & $1646.54$ & $1615.32$ & $5933.33$ \\
\hline 
\end{tabular} }
\end{center}
\caption{Constraints at 2$\sigma$ C.L for the $\Lambda$CDM model. For all baselines, we provide results with and without BBN constraints. Also, we include the constraints for two SNIa baselines.}
\label{tabresultslcdm}
\end{table*}


\begin{table*}[htpb!]
\begin{center}
\renewcommand{\arraystretch}{0.6}
\resizebox{\textwidth}{!}{
\begin{tabular}{l |c |c |c |c |c |c c c c c c c c c c }
\hline
\textbf{Parameter} & \textbf{DESI+BBN} & \textbf{DESI+CMB\footnote{CMB Distance priors}} & \textbf{DESI+CC } & \textbf{\thead{DESI+CMB\\{+CC+$\text{PN}^{+}$}}} & \textbf{\thead{DESI+CC\\+PN$^{+}$}} & \textbf{\thead{DESI+CC\\+SNYR5}}  \\ 
\hline\hline 
$ H_0  $ [km/s/Mpc]& $ 68.5 \pm 6.0 $ & $69.8\pm2.5$ & $70.1^{+7.2}_{-6.9}$ & $69.1^{+1.5}_{-1.3}$ & $72.4\pm2.2 $ &  $66.5^{+6.1}_{-6.4}$ \\
$ \Omega_\mathrm{cdm} $ & $ 0.249^{+0.031}_{-0.034}  $ & $0.246\pm0.017$ & $0.243^{+0.042}_{-0.046}$ & $0.248^{+0.013}_{-0.012}$ & $0.229^{+0.059}_{-0.079} $ & $0.251^{+0.053}_{-0.057}$ \\
$ p_1  $ & $ 0.04^{+0.45}_{-0.48} $ & $-0.12^{+0.21}_{-0.23}$ & $0.08^{+0.43}_{-0.47}$ & $-0.03^{+0.11}_{-0.12}$ & $0.32\pm0.22$ & $0.29\pm0.20$ \\ 
$ w_b  $ & $ 0.02218^{+0.00077}_{-0.00076} $ & $0.02260^{+0.00033}_{-0.00033}$ & $0.027^{+0.013}_{-0.012} $ & $0.02271^{+0.00039}_{-0.00041}$ & $0.0377^{+0.011}_{-0.0099} $ & $0.026^{+0.013}_{-0.012}$ \\
$ \Omega_\mathrm{m} $ & $0.298^{+0.029}_{-0.028} $ & $0.292^{+0.021}_{-0.020}$ & $0.297^{+0.029}_{-0.029}$ & $0.295^{+0.015}_{-0.014}$  & $0.301^{+0.049}_{-0.068} $ & $0.308\pm0.041$ \\ 
$ r_\mathrm{d}  $ [Mpc] & $ 150.6^{+9.9}_{-8.0} $ & $147.12^{+0.79}_{-0.76} $ & $144^{+14}_{-13}$ & $147.24^{+0.88}_{-1.0}$ & $145.1^{+5.1}_{-4.6}$ &  $147^{+15}_{-14} $ \\
$ M $ & $ - $ & $ - $ & $ - $ & $-19.387\pm0.040$ & $ -19.261^{+0.065}_{-0.067}$ &  $-19.51^{+0.20}_{-0.22}$ \\ 
\hline \hline 
$\chi^2_\mathrm{min}$ & $8.83$ & $11.57$ & $23.86$ & $1646.15$ & $1607.846$ & $5926.77$ \\
 $\ln \mathcal B_{ij}$ & $0.102$ & $-0.221$ & $-1.68$ & $-2.46$ & $-2.51$ & $2.66$ \\
\hline
\end{tabular} }
\end{center}
\caption{Constraints at 2$\sigma$ C.L for the $f_1$ model. For all baselines, we provide results with and without BBN constraints. Also, we include the constraints for two SNIa baselines.}
\label{tabresults1}
\end{table*}


\begin{table*}[htpb!]
\begin{center}
\renewcommand{\arraystretch}{0.6}
\resizebox{\textwidth}{!}{
\begin{tabular}{l |c |c |c |c |c |c c c c c c c c c c }
\hline
\textbf{Parameter} & \textbf{DESI+BBN} & \textbf{DESI+CMB\footnote{CMB Distance priors.}} & \textbf{DESI+CC } & \textbf{\thead{DESI+CMB\\{+CC+$\text{PN}^{+}$}}} & \textbf{\thead{DESI+CC\\+PN$^{+}$}} & \textbf{\thead{DESI+CC\\+SNYR5}}  \\ 
\hline\hline 
$ H_0  $ [km/s/Mpc]& $ 68.7^{+2.9}_{-6.2}$ & $67.9^{+1.6}_{-2.1}$ & $69.1^{+7.2}_{-7.1}$ & $68.5\pm1.1$ & $72.3\pm2.0$ & $66.5^{+6.1}_{-6.3} $  \\
$ \Omega_\mathrm{cdm} $ & $ 0.258^{+0.030}_{-0.027} $ & $0.257^{+0.018}_{-0.015}$ & $0.254^{+0.033}_{-0.033}$ & $0.251\pm0.011$  & $0.245^{+0.041}_{-0.039}$ & $0.309^{+0.037}_{-0.043}$\\ 
$ 1/p_2  $ & $ 0.28^{+0.38}_{-0.34}  $ & $0.15^{+0.18}_{-0.17} $ & $0.29^{+0.37}_{-0.34} $ & $0.16^{+0.14}_{-0.19}$ & $0.38^{+0.20}_{-0.22} $ & $0.21^{+0.21}_{-0.23}$  \\
$ w_b $ & $ 0.02219^{+0.00076}_{-0.00076} $ & $0.02260^{+0.00034}_{-0.00034}$ & $0.026^{+0.011}_{-0.011}$ & $0.02277^{+0.00044}_{-0.00057}$ & $0.0336^{+0.0073}_{-0.0076}$ & $0.0215^{+0.0096}_{-0.0090}$ \\ 
$ \Omega_\mathrm{m} $ & $ 0.308^{+0.032}_{-0.028}  $ & $0.306^{+0.021}_{-0.017}$ & $0.308^{+0.030}_{-0.028} $ & $ 0.300\pm0.014$  & $0.309^{+0.034}_{-0.032}$ & $0.357^{+0.031}_{-0.036}$ \\ 
$ r_\mathrm{d}  $ [Mpc] & $ 150.7^{+6.2}_{-5.4} $ & $147.30^{+0.74}_{-0.73}$ & $144^{+14}_{-13}$ & $147.32^{+0.98}_{-0.94}$ & $145.8^{+4.7}_{-4.3} $ & $145^{+15}_{-14}$ \\
$ M $ & $ - $ & $ - $ & $ - $ & $-19.396^{+0.033}_{-0.044}$ & $-19.258^{+0.062}_{-0.066}$ & $-19.52^{+0.20}_{-0.21}$ \\ 
\hline \hline 
$\chi^2_\mathrm{min}$ & $8.64$ & $12.47$ & $23.749$ & $1644.85$ & $1606.53$ & $5932.84$ \\
 $\ln \mathcal B_{ij}$ & $0.839$ & $1.03$ & $-0.321$ & $1.37$ & $2.43$ &$4.73$ \\
\hline
\end{tabular} }
\end{center}
\caption{Constraints at 2$\sigma$ C.L for the $f_2$ model. For all baselines, we provide results with and without BBN constraints. Also, we include the constraints for two SNIa baselines.}
\label{tabresults2}
\end{table*}


To derive our extended cosmology, we start with the 
the action \cite{Ferraro:2006jd,Linder:2010py,RezaeiAkbarieh:2018ijw}: 
\begin{equation}
    \mathcal{S}_{\mathcal{F}(T)}^{} =  \frac{1}{2\kappa^2}\int \mathrm{d}^4 x\; e\left[-T + \mathcal{F}(T)\right] + \int \mathrm{d}^4 x\; e\mathcal{L}_{\text{m}}\,,
\end{equation}
where $\kappa^2=8\pi G$ and the tetrad determinant is calculated as $e=\det\left(\udt{e}{a}{\mu}\right)=\sqrt{-g}$.
When $\mathcal{F}(T) \rightarrow 0$, we recover the concordance $\Lambda$CDM model. As it is oftentimes, we consider a flat homogeneous and isotropic geometry as $\udt{e}{A}{\mu} = \text{diag}\left(1,\,a(t),\,a(t),\,a(t)\right)\,$ \cite{Krssak:2015oua,Tamanini:2012hg}, where $a(t)$ is the scale factor. Using the relationship between the metric and the tetrad $g_{\mu\nu} = \udt{e}{A}{\mu}\udt{e}{B}{\nu}\eta_{AB}\,$, we can write the flat Friedmann--Lema\^{i}tre--Robertson--Walker (FLRW) metric as
\begin{equation}
     \mathrm{d}s^2 = \mathrm{d}t^2 - a^2(t) \left(\mathrm{d}x^2+\mathrm{d}y^2+\mathrm{d}z^2\right),
\end{equation}
with $H=\dot{a}/a$. Subsequently, we can derive the Friedmann equations:
\begin{eqnarray}
    H^2 + \frac{T}{3}\mathcal{F}_T - \frac{\mathcal{F}}{6} &= \frac{\kappa^2}{3}\rho\,\\
    \dot{H}\left(1 - \mathcal{F}_T - 2T\mathcal{F}_{TT}\right) &= -\frac{\kappa^2}{2} \left(\rho + p \right),
\end{eqnarray}
where $\rho$ and $p$, are the energy density and pressure, respectively. We selected $f(T)$ cases where it is possible to reproduce naturally a late-time cosmic acceleration:
\begin{itemize}
\item \textit{Power Law Model ($f_1$)}\cite{Bengochea:2008gz}.  -- 
This model is of the form: $f_1 (T) = \left(-T\right)^{p_1}\,,$
where $p_1$ is a constant. We can recover $\Lambda$CDM model when $p_1 = 0$. Otherwise, if $p_1 = 1$, the extra term gives a re-scaled gravitational constant related to the GR limit. Furthermore, when $p_1 < 1$ gives an accelerating universe. To compare the new constraints for this model using DESI 2024, in \citep{Xu:2018npu} was considered BAO measurements from Two-Degree Field Galaxy Redshift Survey (2dFGRS) and SDSS DR7, where it was found that $H_0 = 69.4 \pm 0.8$[km/s/Mpc], $\Omega_m = 0.298 \pm 0.07$ and $p_1 = -0.10^{+0.09}_{-0.07}$, where it is suggested that a deviation from the $\Lambda$CDM model is present in the datasets. 

\item \textit{Linder Model ($f_2$)}\cite{Linder:2010py}. -- 
This model is described as: $f_2 (T) =  T_0 (1 - e^{[-p_2\sqrt{T/T_0}]})\,,$ where $p_2$ is a constant and $T_0 = T\vert_{t=t_0} = -6H_0^2$. Notice that this model recovers $\Lambda$CDM in the limit $p_2 \rightarrow +\infty$. As in the latter case, this model was tested using BAO from 2dFGRS and SDSS obtaining $H_0 = 69.6 \pm 0.9$[km/s/Mpc], $\Omega_m = 0.296 \pm 0.07$ and $1/p_2 = 0.13^{+0.09}_{-0.11}$ \citep{Xu:2018npu}, again, denoting an interesting deviation, yet small deviation from the standard cosmological model. 
\end{itemize}

\begin{figure*}
    \centering
    \includegraphics[width=0.32\textwidth]{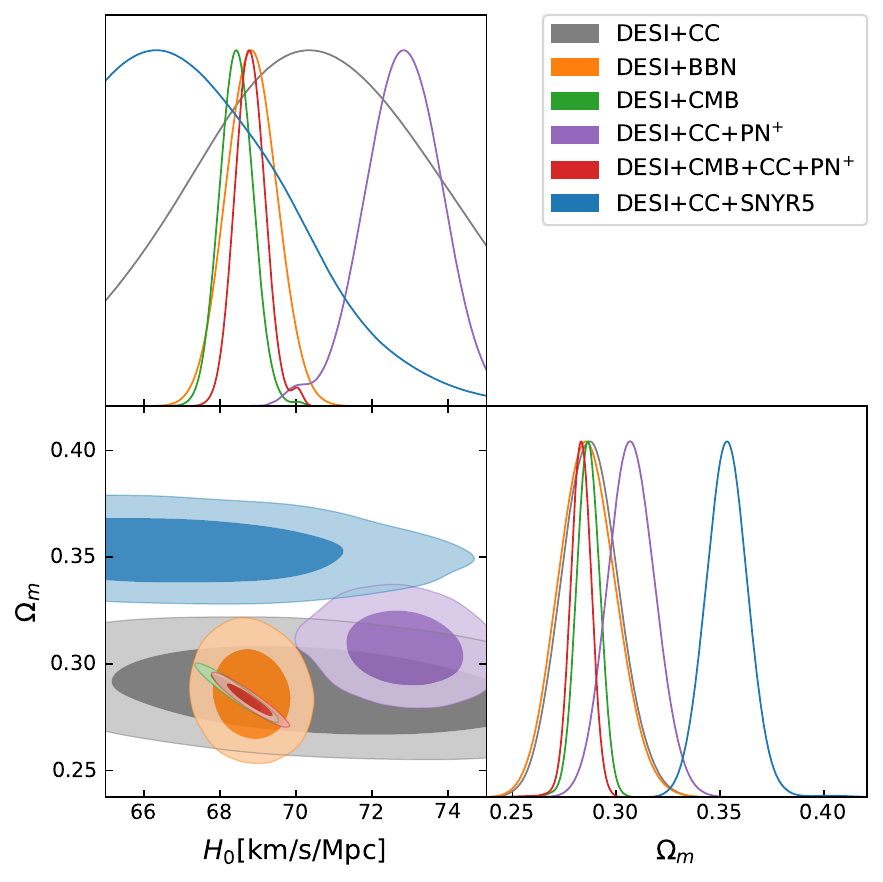}
     \includegraphics[width=0.32\textwidth]{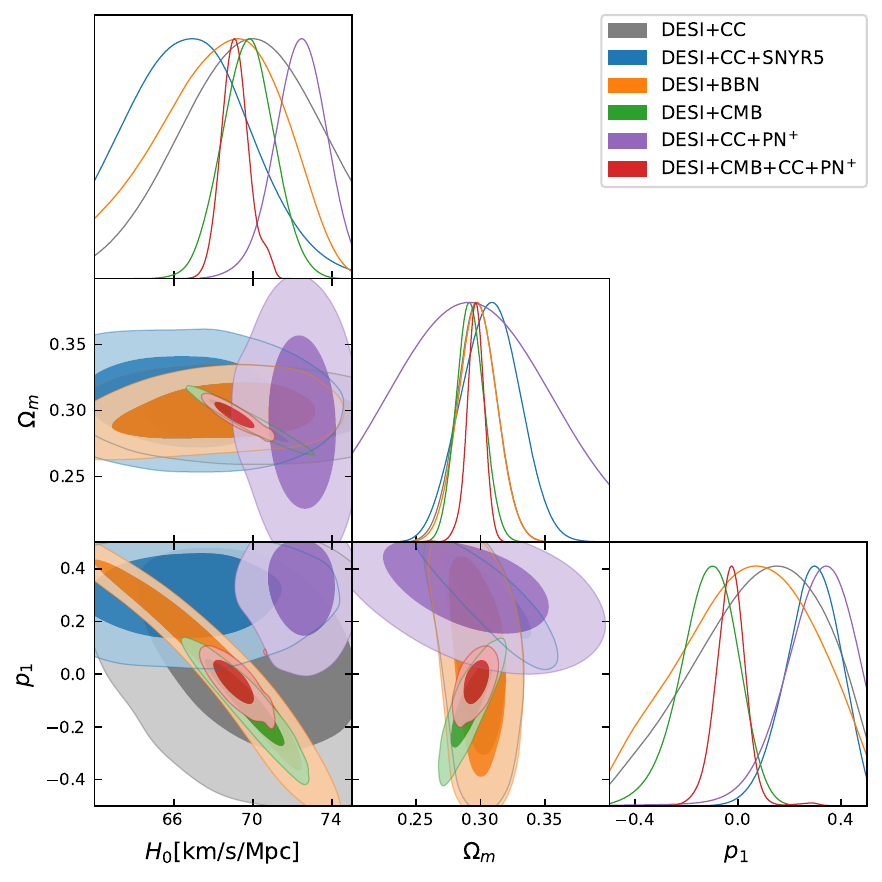}
         \includegraphics[width=0.32\textwidth]{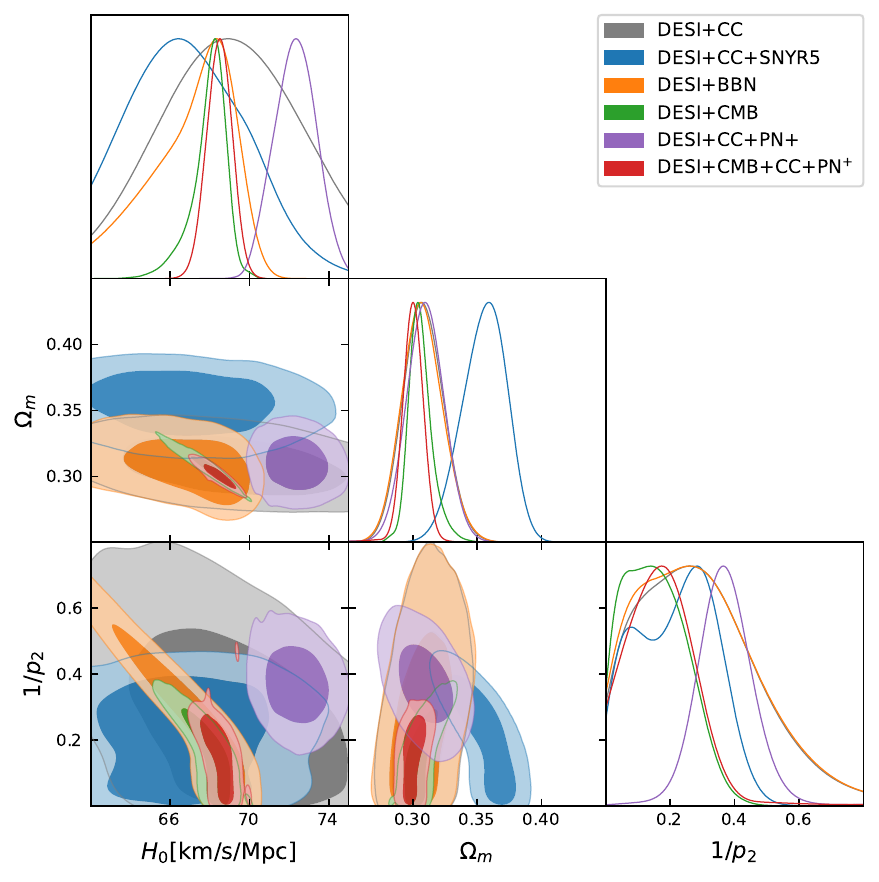}
\caption{1$\sigma$ and 2$\sigma$ Confidence Levels (C.L) and posterior distributions including $H_0$ and $\Omega_{m,0}$. The baselines are indicated in colours for each case. \textit{Left:} For the $\Lambda$CDM model. \textit{Middle:} For the power law model $f_1$. \textit{Right:} For the Linder model $f_2$.}
    \label{fig:lcdm_all}
\end{figure*}



We implement each of $f(T)$ cosmological model described and test them using the constraining parameters method through MCMC analysis using \texttt{emcee} \footnote{\href{https://emcee.readthedocs.io/en/stable/}{emcee.readthedocs.io}} for the cosmology and the baselines with the extract of constraints using \texttt{GetDist}\footnote{\href{https://getdist.readthedocs.io/en/latest/}{getdist.readthedocs.io}}. Additionally, we assume flat priors on the set of \{$\Omega_b h^2$, $\Omega_c$, $H_0$\}. The baselines considered in our analysis are:
\textit{(a)} \textbf{DESI(BAO)} release obtained from observations of galaxies and quasars~\cite{DESI:2024uvr}, and Lyman-$\alpha$~\cite{DESI:2024lzq} measurements. These trasers are described through the transverse comoving distance $D_{\mathrm{M}}/r_{\mathrm{d}}$, the angle-averaged distance $D_{\mathrm{V}}/r_{\mathrm{d}}$, where  $r_{\mathrm{d}}$ is the comoving sound horizon at the drag epoch, and the Hubble horizon $D_{\mathrm{H}}/r_{\mathrm{d}}$. 
\textit{(b)} \textbf{CMB} Planck-2018 distant priors, which provide information on three parameters: the shift parameter $R$ that measures the peak spacing of the temperature in the CMB spectrum, the acoustic scale $l_\mathrm{A}$ from we can measure the temperature in the transverse direction, and finally the combination $\Omega_b h^2$ \cite{Chen:2018dbv}.
\textit{(c)} Cosmic Chronometers \textbf{CC}, which are measurements of $H(z)$ from the relative ages of passively-evolving galaxies \cite{Jimenez:2001gg}. We conservatively use the galactic spectra to obtain $\mathrm{d}t/\mathrm{d}z$ \citep{Moresco:2016mzx}. The final sample contains 31 data points up to $z \sim 2$ with the covariance matrix generated given in \citep{Moresco:2020fbm}. 
\textit{(d)} \textbf{PN${}^{+}$} Pantheon-plus catalog~\cite{Brout:2022vxf}, with SH0ES Cepheid host distances calibrators~\cite{Riess:2021jrx}, and
\textit{(e)} \textbf{DES-SN5YR} Type Ia supernovae measured during the full 5-years of DES Supernova Program, which includes 1635 SNIa in the redshift range $0.10<z<1.13$ \cite{DES:2024tys}.

We divide our analysis into these baselines since $H_0$ and $r_d$ are degenerate in the DESI BAO release. Due to this degeneracy, we will test the set in different schemes: 
(\textit{i}) Using the combination of parameters $\Omega_m$ and $r_d h$ in Mpc to avoid the degeneracy between $h = H_0/100 \text{[km/s/Mpc]}$ and $r_d$. This yields the results with the 95\% confidence intervals for the $\Lambda$CDM model: 
\begin{equation*}
    \left\{\begin{matrix}
        \Omega_m = 0.286^{+0.028}_{-0.026}, \\ 
        r_d h = 102.6\pm 2.5 \text{ [Mpc]},
    \end{matrix} \right. \text{DESI(BAO)}
\end{equation*}
which is in $2\sigma$ interval from the results reported by the DESI collaboration \citep{desicollaboration2024desi}. Meanwhile, for the $f_1(T)$ model:  
\begin{equation*}
    \left\{\begin{matrix}
        \Omega_m = 0.282^{+0.031}_{-0.033}, \\ 
        r_d h = 102.0\pm 4.2 \text{ [Mpc]}, \\
        p_1 = 0.06^{+0.44}_{-0.48}.
    \end{matrix} \right. \text{DESI(BAO)}
\end{equation*}
we can notice that the data prefers a slightly lower fractional matter density with a similar product $r_dh$ and that the free parameter for the power-law model $p_1$ is within $2\sigma$ region. This recovers $\Lambda$CDM with a minor positive deviation. In this case, the $f_1(T)$ model contains a Bayes factor of $\ln \mathcal B_{ij} = +1.34$, which indicates a slight preference for the $\Lambda$CDM. For the $f_2(T)$ model:
\begin{equation*}
    \left\{\begin{matrix}
        \Omega_m = 0.307^{+0.041}_{-0.039}, \\ 
        r_d h = 100.6^{+4.5}_{-5.1} \text{ [Mpc]}, \\
        1/p_2 = 0.29^{+0.37}_{-0.33}. \\
    \end{matrix} \right. \text{DESI(BAO)}
\end{equation*}
Contrary to the previous model, here the fractional matter exhibits an increase and a diminution in the $r_dh$ parameter. For the free Linder model parameter $1/p_2$ this dataset alone recovers the $\Lambda$CDM model as $1/p_2 \to 0$ in $2\sigma$ limit. The Bayes factor must be compared to the tested one for the $\Lambda$CDM model which results in $\ln \mathcal B_{ij} = +0.33$, this favours the standard cosmological model.
(\textit{ii}) By using a prior on $r_d$ from Planck 2018 \citep{Planck:2018vyg} of $r_d = 147.09 \pm 0.87$[Mpc] it is possible to break the degeneracy with $H_0$. The results within 95\% confidence interval for the $\Lambda$CDM model are: 
\begin{equation*}
    \left\{\begin{matrix}
        H_0 = 69.7\pm 1.7 \text{[km/s/Mpc]}, \\
        \Omega_m = 0.286\pm 0.029, 
    \end{matrix} \right. \text{DESI +}r_d ~\text{CMB Planck}
\end{equation*}
Remarkably interesting, these results have a high $H_0$ value even though we are using a $r_d$ from the Planck estimations. For the $f_1(T)$ model, the parameters are: 

\begin{equation*}
    \left\{\begin{matrix}
        H_0 = 69.4^{+3.0}_{-2.9} \text{[km/s/Mpc]}, \\
        \Omega_m = 0.281^{+0.031}_{-0.034},\\
        p_1 = 0.05^{+0.45}_{-0.48}, 
    \end{matrix} \right. \text{DESI +}r_d ~\text{CMB Planck}
\end{equation*}
where the $H_0$ value shows a compatibility in $2\sigma$ with the value obtained by the SH0ES collaboration \citep{Riess:2021jrx}. However, it is important to note that this result is originated due to the increased error bar in the parameter. This model returns a confirmation of $\Lambda$CDM for the $p_1$ value with a significant error bars, probably because this dataset alone can not constrain the parameter solely. In this case the Bayes factor $\ln \mathcal B_{ij} = -0.23$ which suggests that using a $r_d$ prior to the dataset and the power-law model is slightly favoured over the cosmological standard model. For the $f_2(T)$ model the results are:
\begin{equation*}
    \left\{\begin{matrix}
        H_0 = 68.3^{+3.0}_{-3.5} \text{[km/s/Mpc]}, \\
        \Omega_m = 0.306^{+0.032}_{-0.029},\\
        1/p_2 = 0.29^{+0.36}_{-0.32}, 
    \end{matrix} \right. \text{DESI +}r_d ~\text{CMB Planck}
\end{equation*}
that, similarly to the previous model, confirm $\Lambda$CDM at $2\sigma$ level. In this case, the value of $\Omega_m$ presents a higher value that is a tendency using this specific model. This model presents a Bayes factor $\ln \mathcal{B}_{ij} = -0.19$ that, again, suggests that the $f_2(T)$ model is preferred over $\Lambda$CDM. However, it is important to notice that in each of these models the value of the evidence between the $f(T)$ models and the standard cosmology is not strong enough. 
(\textit{iii}) Using a prior on $w_b =\Omega_b h^2$ using the results of BBN presented in \citep{desicollaboration2024desi} of $w_b = 0.02218 \pm 0.00055$ to break the degeneracy. In this case, we calculate $r_d$ as a derived parameter. This analysis is presented in Tables \ref{tabresultslcdm}, \ref{tabresults1}, \ref{tabresults2} including DESI + BBN. 
(\textit{iv}) Finally, since the uncertainties on this release are substantial we will consider other datasets without the necessity to introduce a prior on $w_b$ as the baselines are sufficient enough to constraint the cosmological parameters. These results are reported in Tables \ref{tabresultslcdm}, \ref{tabresults1}, \ref{tabresults2} in combination with other baselines including DESI BNN measurements.


\bigskip

In conclusion, $f(T)$ cosmologies constrained by new BAO measurements from DESI 2024 (and another dataset that allow us to constraint $H_0$ as the $r_d$ prior from the CMB Planck, BBN, Pantheon+ or CMB Distance Priors) can be a good alternative to explain the current $H_0$ tension as the results using this dataset combinations
show an improvement in the alleviation on the $H_0$ value closer to the SH0ES collaboration. Furthermore, it is important to notice that the mentioned combinations of DESI BAO with other datasets such as CC, Pantheon+, and even the CMB Distance Priors, the statistics show a slight preference for the $f_1$ model. This preference is in addition to the aforementioned advantage of alleviating the Hubble tension. 

New analyses will be conducted using the data released in the coming months, employing these extended gravity models. Finally, this result from DESI BAO 2024 measurements could be a hint that the cosmological tension needs new physics to be solved.


\begin{acknowledgments}
\bigskip
\noindent 
\textit{---} \\
\textit{Acknowledgments.-}
CE-R is supported by the CONACyT Network Project No. 376127 and acknowledges the Royal Astronomical Society as Fellow FRAS 10147. 
RS is supported by the CONACyT National Grant.
This research has been carried out using computational facilities procured through the Cosmostatistics National Group ICN UNAM project.
This article is based upon work from COST Action CA21136 Addressing observational tensions in cosmology with systematics and fundamental physics (CosmoVerse) supported by COST (European Cooperation in Science and Technology).
\end{acknowledgments}
\bibliography{ft-references}
\end{document}